\documentstyle[prd,aps,preprint,epsfig]{revtex}
\newfont\fiverm{cmr5}
\input prepictex
\input pictex
\input postpictex

\begin{document}

\newcommand{\TeV}{\,{\rm TeV}}
\newcommand{\GeV}{\,{\rm GeV}}
\newcommand{\MeV}{\,{\rm MeV}}
\newcommand{\keV}{\,{\rm keV}}
\newcommand{\eV}{\,{\rm eV}}
\def\ap{\approx}
\def\bea{\begin{eqnarray}}
\def\eea{\end{eqnarray}}
\def\beqar{\begin{eqnarray}}
\def\eeqar{\end{eqnarray}}
\def\ler{\lesssim}
\def\gtr{\gtrsim}
\def\beq{\begin{equation}}
\def\eeq{\end{equation}}
\def\haf{\frac{1}{2}}
\def\plb#1#2#3#4{#1, Phys. Lett. {\bf #2B} (#4) #3}
\def\plbb#1#2#3#4{#1 Phys. Lett. {\bf #2B} (#4) #3}
\def\npb#1#2#3#4{#1, Nucl. Phys. {\bf B#2} (#4) #3}
\def\prd#1#2#3#4{#1, Phys. Rev. {\bf D#2} (#4) #3}
\def\prl#1#2#3#4{#1, Phys. Rev. Lett. {\bf #2} (#4) #3}
\def\mpl#1#2#3#4{#1, Mod. Phys. Lett. {\bf A#2} (#4) #3}
\def\rep#1#2#3#4{#1, Phys. Rep. {\bf #2} (#4) #3}
\def\lpp{\lambda''}
\def\ccg{\cal G}
\def\slash#1{#1\!\!\!\!\!/}
\def\rpv{\slash{R_p}}

%%%%%%%%%%%%%%%%%%%%% Main Text %%%%%%%%%%%%%%%%
\setcounter{page}{1}
%\renewcommend{\arraystretch}{1.3}
\draft
%\widetext
\preprint{KAIST-TH 01/02}

\title{Atmospheric and Solar Neutrino Masses and Abelian Flavor Symmetry}

\author{Kiwoon Choi}

\address{Department of Physics,
Korea Advanced Institute of Science and Technology\\
        Taejon 305-701, Korea}

%\date{\today}

\tighten

\maketitle

\begin{abstract}
Recent atmospheric and solar neutrino experiments suggest  
that neutrinos have small but nonzero masses.
They further suggest that
mass eigenvalues have certain degree of hierarchical structures,
and also some mixing angles are near-maximal while the others are small.
We first survey possible explanations
for the smallness of neutrino masses.
We then discuss some models
in which the hierarchical pattern of neutrino
masses and mixing angles arises as a consequence of 
$U(1)$ flavor symmetries which would
explain also the hierarchical quark and charged lepton masses.
  
\end{abstract}

%\pacs{PACS number(s):}

%\begin{multicols}{2}

\section{Introduction}

Atmospheric and solar neutrino experiments have suggested  for
a long time that neutrinos oscillate into different flavors, thereby
have nonzero masses \cite{valle}.
In particular, the recent Super-Kamiokande data strongly
indicates that the observed deficit of atmospheric muon
neutrinos is due to the near-maximal $\nu_{\mu}\rightarrow
\nu_{\tau}$ oscillation \cite{atm}. Solar neutrino results
including those of Super-Kamiokande, Homestake, SAGE and GALLEX
provide also  strong observational basis for $\nu_e\rightarrow\nu_{\mu}$ or 
$\nu_{\tau}$ oscillation \cite{sol}.

The minimal framework to accomodate the atmospheric and 
solar neutrino anomalies is to introduce small but nonzero
masses of the three known neutrino species.
In the basis in which the charged current
weak interactions are flavor-diagonal,
the relevant piece of low  energy effective lagrangian
is given by
\beq
\overline{e_L} M^e e_R +
  g W^{-\mu}\overline{e_L}\gamma_\mu \nu_L 
  + \overline{(\nu_L)^c} M^\nu \nu_L \, ,
\eeq
where the $3\times 3$ mass matrices $M^e$ and $M^\nu$ are not diagonal
in general. 
Diagonalizing $M^e$ and $M^{\nu}$, 
\bea
&& (U^e)^{\dagger} M^e V^e = D^e=
{\rm diag} \, (m_e, m_{\mu},
m_{\tau}), \nonumber \\
&& (U^\nu)^T M^{\nu} U^\nu = D^{\nu}=
{\rm diag} \, (m_1, m_2, m_3),
\eea
one finds the effective lagrangian written in terms of the mass eigenstates
\beq
\overline{e_L} D^{e} e_R +
  W^{-\mu}\overline{e_L}\gamma_\mu U \nu_L 
  + \overline{(\nu_L)^c} D^\nu \nu_L\,,
\eeq
where the MNS lepton mixing matrix is given by
\beq
\label{mns}
U = (U^e)^{\dagger} U^\nu.
\eeq
Upon ignoring CP-violating phases, $U$ can be parametrized as
\bea
\label{para}
U &=& \pmatrix{ 1 & 0 & 0 \cr 0 & c_{23} & s_{23} \cr 0 & -s_{23} & c_{23} }
    \pmatrix{ c_{13} & 0 & s_{13} \cr 0 & 1 & 0 \cr -s_{13} & 0 & c_{13} }
    \pmatrix{ c_{12} & s_{12} & 0 \cr -s_{12} & c_{12} & 0 \cr 0 & 0 & 1 }
\nonumber \\
&=& \pmatrix{ c_{13}c_{12} & s_{12}c_{13} &
s_{13} \cr -s_{12}c_{23}-s_{23}s_{13}c_{12} & c_{23}c_{12}-s_{23}s_{13}s_{12}
& s_{23}c_{13} \cr s_{23}s_{12}-s_{13}c_{23}c_{12}
& -s_{23}c_{12}-s_{13}s_{12}c_{23} & c_{23}c_{13}}
\eea
where $c_{ij} = \cos \theta_{ij}$ and $s_{ij} = \sin \theta_{ij}$.
Within this parameterization, the mass-square differences for
atmospheric and solar neutrino oscillations can be chosen to be
\beq
\Delta m^2_{\rm atm}=m_3^2-m_2^2,
\quad
\Delta m^2_{\rm sol}=m_2^2-m_1^2,
\eeq
while the mixing angles are given by
\beq
\theta_{\rm atm}=\theta_{23},
\quad 
\theta_{\rm sol}=\theta_{12},
\quad
\theta_{\rm rea}=\theta_{13},
\eeq
where $\theta_{\rm rec}$ describes for instance
the neutrino oscillation $\nu_{\mu}\rightarrow
\nu_e$ in reactor experiments.

The atmospheric neutrino data sugget near-maximal
$\nu_{\mu}\rightarrow \nu_{\tau}$ oscillation \cite{atm} with
\bea
&& \Delta m^2_{\rm atm}\sim 3\times 10^{-3} \, {\rm eV}^2,
\nonumber \\
&& \sin^2 2\theta_{\rm atm}\sim 1.
\eea
As for the solar neutrino anomaly, four different
oscillation scenarios are possible \cite{sol} though the large
mixing angle (LMA)
MSW oscillation is favored by the recent Super-Kamiokande data:
\bea
&& {\rm SMA \, \, MSW}: \Delta m^2_{\rm sol}
\sim 5\times 10^{-6} \, {\rm eV}^2,
\quad \sin^2 2\theta_{\rm sol}\sim 5\times 10^{-3},
\nonumber \\
&&
{\rm LMA \, \, MSW}: \Delta m^2_{\rm sol}
\sim 2\times 10^{-5} \, {\rm eV}^2,
\quad \sin^2 2\theta_{\rm sol}\sim 0.8,
\nonumber \\
&&
{\rm LOW \, \, MSW}: \Delta m^2_{\rm sol}
\sim 10^{-7} \, {\rm eV}^2,
\quad \sin^2 2\theta_{\rm sol}\sim 1,
\nonumber \\
&&
{\rm LMA \, \, VAC}: \, \Delta m^2_{\rm sol}\sim
10^{-10} \, {\rm eV}^2,
\quad \sin^2 2\theta_{\rm sol}\sim 0.7 \, .
\eea 
There is in fact an important constraint from reactor
experiments, e.g. CHOOZ \cite{CHOOZ}, indicating no $\nu_{\mu}$
oscillation into $\nu_e$, thereby leading to
\beq
4U_{e3}^2\approx \sin^2 2\theta_{13} \lesssim 0.15
\eeq

Putting the atmospheric and solar neutrino data together
while taking into account the CHOOZ constraint, one can
consider the following three patterns of neutrino masses and
mixing angles:

I.  Bi-maximal mixing with LMA MSW solar neutrino oscillation:
\bea
&& m_2/m_3 \, \sim \, \lambda \, \, {\rm or} \, \, \lambda^2 \, ,\nonumber \\
&& (\, |s_{23}|,\, |s_{12}| , \, |s_{13}| \,)
\, \sim \, (\frac{1}{\sqrt{2}}, \, \frac{1}{\sqrt{2}}, \, \lambda^k)
 \, ,
\eea

II. Bi-maximal mixing with LOW MSW or LMA VAC solar neutrino oscillation:
\bea
&& m_2/m_3 \, \sim \,  \lambda^4 \, \, {\rm or} \, \, \lambda^5 \, ,
\nonumber \\
&& (\, |s_{23}|, \, |s_{12}|, \, |s_{13}| \,)
\, \sim \,  (\frac{1}{\sqrt{2}},\, \frac{1}{\sqrt{2}},\, \lambda^k)
 \, ,
\eea

III.  Single-maximal mixing with SMA MSW solar neutrino oscillation:
\bea
&& m_2/m_3 \, \sim \,  \lambda^2  \, ,
\nonumber \\
&& (\, |s_{23}|, \, |s_{12}|,\, |s_{13}| \,)
\sim (\frac{1}{\sqrt{2}}, \, \lambda^2, \, \lambda^k)
 \, ,
\eea
where $\lambda\equiv \sin \theta_C \sim 0.2$ 
for the Cabbibo angle $\theta_C$ and
\beq
m_3 \, \sim \, 5\times 10^{-2} \, \, {\rm eV} \, ,
\quad \quad k\geq 1
\eeq
in all cases.
These neutrino results can be compared 
with the following quark and charged lepton masses and mixing angles:
\bea
&& (\, m_t, m_c, m_u \,) \, \sim \,
 180 \, (\, 1, \, \lambda^4, \, \lambda^8) \quad {\rm GeV},
\nonumber \\
&& (\, m_b, m_s, m_d \,) \, \sim \, 4  \, (\, 1, \, \lambda^2, \, \lambda^4) 
\quad {\rm GeV},
\nonumber \\
&& (\, m_{\tau}, m_{\mu}, m_e \,) \,  \sim \, 1.8 \, 
(\, 1, \, \lambda^2, \, \lambda^5) 
\quad {\rm GeV},
\nonumber \\
&&
(\, \sin {\phi}_{23}, \, \sin {\phi}_{12},
\, \sin {\phi}_{13} \,)\sim
(\, \lambda^2, \, \lambda, \, \lambda^3),
\eea 
where $\phi_{ij}$ denote the mixing angles of the 
Cabbibo-Kobayashi-Maskawa matrix parametrized as  
(\ref{para}).

The above neutrino masses and mixing angles involve some
small numbers. Obviously, the most distinctive 
one is $m_3/m_{\tau}\sim 3\times 10^{-10}$, leading to the
old question ``why neutrinos
are so light compared to their charged lepton counterparts ?''
Though not extremely small as $m_3/m_{\tau}$,
there are other small numbers like $m_2/m_3$,
$1-\sin^2 2\theta$ for near-maximal mixing
and $\sin^2 2\theta$ for small mixing, which may require
some explanations.
For instance, for the scenarios I and II, one may wonder
how near maximal $\theta_{23}$ and small $m_2/m_3$
can be simultaneously obtained and also how $\theta_{13}$
can be made to be small 
while keeping $\theta_{23}$ and $\theta_{12}$ near maximal. 
Similarly, for the scenario III, one can ask what 
would be the flavor structure
yielding small $m_2/m_3\sim \lambda^2$
and $\theta_{12}\sim \lambda^2$, but near maximal $\theta_{23}$.
In this talk, we first survey possible explanations for
the smallness of neutrino masses, and then discuss some models
in which the hierarchical patterns of neutrino masses 
and mixing angles arise as a consequence $U(1)$ flavor symmetries
\cite{u1,otheru1} 
which would explain also the hierarchical quark and charged lepton
masses.

\section{Why neutrinos are so light?}

Here we discuss four possible mechanisms which would 
suppress the resulting neutrino mass.
These mechanisms are not orthogonal to each other, 
so one can take more than one mechanism  in order to make
the neutrino mass small enough.

\medskip

{\bf A. \,   Seesaw-type mechanism}

\medskip

In seesaw-type model, neutrinos are light since their masses
are induced by the exchange of superheavy particles.
At low energies, the effects of such heavy particles are described
by  the operator
\beq
\label{operator}
\frac{1}{M}LLHH
\eeq
where $L$ and $H$ are the lepton and Higgs doublets, respectively,
and $M$ denotes the mass scale of
the exchanged heavy particle.
This gives a neutrino mass
\beq 
m_{\nu}\sim \frac{\langle H\rangle\langle H\rangle}{M}
\sim 5\times 10^{-2} \left(\frac{10^{15} \, {\rm GeV}}{M}
\right) \, {\rm  eV}
\eeq
which can be as small as the atmospheric neutrino mass
for $M\sim 10^{15}$ GeV.
There are two different ways to generate the above $d=5$ operator.
One is the exchange of superheavy  singlet neutrino
\cite{seesaw} (Fig. 1) which corresponds to the conventional seesaw mechanism,
 and the other is the exchange of superheavy triplet Higgs boson
\cite{triplet} (Fig. 2). 
For the case of singlet neutrino exchange, 
the underlying lagrangian includes
\beq
hHLN + M_N NN  +{h.c.},
\eeq
where $N$ is a singlet neutrino with huge Majorana mass $M_N$.
Integrating out $N$ then yields the operator
(\ref{operator}) with $M=M_N/h^2$.
For the case of triplet Higgs exchange \cite{triplet,santa},
one starts from 
\beq
h T LL -\frac{1}{2}M_{T}^2TT^*-
M^{\prime}_{T}T H^*H^* +{h.c.},
\eeq
where $T$  is a Higgs triplet with huge mass $M_T$.
Again integrating out $T$ leads to (\ref{operator}) with
$M=M_T^2/h M^{\prime}_T$.
The seesaw-type mechanism is perhaps
the simplest way to get small
neutrino mass. However it would be rather difficult to probe
other effects of the involved superheavy particles 
than generating the neutrino mass.

\medskip

{\bf B. \, Frogatt-Nielsen mechanism}

\medskip

Neutrino mass can be small if the couplings which are
responsible for neutrino mass are suppressed by a spontaneously 
broken (gauge) symmetry by means of the Frogatt-Nielsen mechanism \cite{FN}.
As an example, consider  again a model with singlet neutrino
$N$ but now with weak scale $M_N\sim 10^2$ GeV.
Suppose that the operator $HLN$ carries a nonzero integer charge
$n$ of some discrete or continuous (gauge) symmetry $G$ of the model
and $G$ is spontanesly broken by the VEV of a standard model
singlet $\phi$ which has charge $-1$.
Then $HLN$ in the bare action is forbiden,
however there can be higher-dimensional coupling
$\phi^n HLN/M_*^n$ allowed by $G$ where $M_*$ corresponds to the 
fundamental (or UV-cutoff) scale of the model.
This results in the effective Yukawa coupling
\beq
\left(\frac{\langle \phi\rangle}{M_*}\right)^n HLN\equiv
\epsilon^n HLN,
\eeq
and so the neutrino mass
\beq
m_{\nu}\sim \frac{\epsilon^{2n}\langle
H\rangle\langle H\rangle}{M_N}\sim 5\times 10^{-2} 
\left(\frac{\epsilon^{n}}{3\times 10^{-7}}\right)^2\left(
\frac{10^2 \, {\rm GeV}}{M_N}\right) \, \, {\rm eV}.
\eeq
By choosing appropriate values of $n$ and $\epsilon$, one
can easily accomodate the atmospheric neutrino mass
even when the singlet neutrino  has an weak scale
mass.

In supersymmetric models, one can implement the Frogatt-Nielsen
mechanism for small neutrino mass without introducing
any singlet neutrino.
As an example, consider a supersymmetric model 
with $U(1)$ flavor symmetry whose symmetry breaking order
parameter $\epsilon\sim \lambda$ (Cabbibo angle).
The $U(1)$ charges of $H_1H_2$ and $LH_2$ are assumed to be
$-1$ and $-n$, respectively, 
where $H_{1,2}$ and $L$ denote the 
two Higgs doublets and the lepton doublet superfields in the MSSM.
Then the supergravity 
K\"ahler potential can contain
\beq
K= \frac{\phi^*}{M_*}H_1H_2 +\left(\frac{\phi^{*}}{M_*}
\right)^nLH_2,
\eeq
while the holomorphy and $U(1)$ do not allow
the supergravity superpotential contain a term 
like $\phi^mH_1H_2$ or $\phi^mLH_2$.
After the spontaneous breaking of supersymmetry and 
also of $U(1)$, this K\"ahler potential
gives rise to the $\mu$-type terms
in the effective superpotential:
$$
W_{eff}=\mu H_1H_2+\mu^{\prime}LH_2,
$$
where
$\mu\sim \epsilon^* m_{3/2}\sim
\lambda m_{3/2}$ and $\mu^{\prime}\sim \epsilon^{*n} m_{3/2}
\sim \lambda^n m_{3/2}$.
Here the first term in $W_{eff}$ is just the
conventional $\mu$-term and the second corresponds to
the bilinear $R$-parity violating term. 
As is well known,  the bilinear $R$-parity violation
leads to the neutrino mass \cite{HS}
\beq
m_{\nu}\sim  \frac{\mu^{\prime 2}\langle H_2\rangle \langle
H_2\rangle}{\mu^2M_{1/2}}\sim
\lambda^{2(n-1)}M_{weak}
\eeq
where the gaugino mass $M_{1/2}$ and $\mu$ are assumed to
have  the weak scale value $M_{weak}$.
This neutrino mass can be of order the atmospheric neutrino
mass if $n=9$, which would be obtained
for instance if the $U(1)$ charges of $H_1$,
$H_2$, $L$ are $4$, $-5$, $-4$, respectively.

In fact, supersymmetric models always contain an
intrincically small symmetry breaking parameter,
i.e. $m_{3/2}/M_{Planck}$ describing the size
of SUSY breaking. 
If $m_{\nu}$ is suppressed by a symmetry $G$ which is
broken by the SUSY breaking dynamics, 
small $m_{\nu}/M_{weak}$  and $m_{3/2}/M_{Planck}$
have a common dynamical origin.
They are then related to each other
by the Frogatt-Nielsen mechanism of $G$, e.g
\beq
\frac{m_{2/3}}{M_{Planck}}\sim \epsilon^k,
\quad
\frac{m_{\nu}}{M_{weak}}\sim \epsilon^l,
\eeq
where $\epsilon$ is the symmetry breaking order
parameter of $G$, and $k$ and $l$ are model-dependent
integers. If $k/l=4/3$ and $m_{3/2}\sim M_{weak}$, 
one obtains $m_{\nu}/M_{weak}\sim 10^{-12}$
which is the correct value for the atmospheric neutrino mass.
$m_{3/2}/M_{Planck}$ and $m_{\nu}/M_{weak}$ is
always related to each other when $G$ is a discrete $R$-symmetry
\cite{choi1} 
which appears quite often in compactified string theory.
It is also possible to relate $m_{\nu}/M_{weak}$
with $m_{3/2}/M_{Planck}$ by means of other type of symmetry
\cite{hall}.

\medskip

{\bf C. \, Radiative generation of neutrino mass}

\medskip

Even when the neutrino mass is zero at tree level, if
the lepton number symmetry is softly broken, there can be
small finite radiative corrections to neutrino mass \cite{radiative}.
The resulting neutrino mass is suppressed by the loop factor
as well as the (potentially) small Yukawa couplings which are 
involved in the loop.
The most typical example is the Zee model with
$$
 {\cal L}=fHLE^c+f^{\prime}S^+LL-AS^+HH^{\prime}+...,
$$
where $H$ and $H^{\prime}$ are $SU(2)$-doublet Higgs fields, 
$S^+$ is a charged $SU(2)$-singlet Higgs field, and $L$ and $E^c$ stand for
the conventional lepton doublet and anti-lepton singlet.
Here we will ignore the flavor indices of couplings for simplicity.
It is then easy to see that a nonzero neutrino mass
is generated at one-loop (Fig. 3), yielding
\beq
m_{\nu}\sim
\frac{f^{\prime}f^2A}{16\pi^2m_S^2}\langle H\rangle
\langle H^{\prime}\rangle,
\eeq
where $m_S$ is the mass of $S^+$.

It is also possible to construct a model in which $m_{\nu}$
is generated at higher loop order \cite{radiative}.
A typical example is given by
$$
{\cal L}=fHLE^c+f^{\prime}S^+LL+f^{\prime\prime}
S^{--}E^cE^c-AS^+S^+S^{--}+...,
$$
where $S^+$ and $S^{--}$ are charged
$SU(2)$-singlet Higgs fields.
In this model,  nonzero $m_{\nu}$ appears at two-loop (Fig. 4), yielding
\beq
m_{\nu}\sim \frac{Af^2f^{\prime2}f^{\prime\prime}}{(16\pi^2)^2m_S^2}
\langle H\rangle\langle H\rangle.
\eeq
Note that even when 
$S^+$ and $S^{--}$ have weak
scale masses, the resulting neutrino mass can be 
as small as the atmospheric 
neutrino mass if the Yukawa couplings $f,f^{\prime},f^{\prime}
\sim 10^{-2}$.

\medskip

{\bf D. \, Localizing singlet neutrino on the hidden brane}

\medskip

Recently it has been noted by Randall and Sundrum (RS) that 
the large hierarchy
between $M_{Planck}$ and $M_{weak}$ can be achieved
by localizing the gravity on a hidden brane \cite{randall}.
In the RS model, the spacetime is given by a slice
of $d=5$ AdS space with two boundaries.
A flat 3-brane with positive tension is sitting on one
of these boundaries at $y=0$, while a negative tension
3-brane is on the other boundary at $y=b$.
Massless $d=4$ graviton mode is localized on the positive
tension brane (the hidden brane)
while the observable standard model fields 
are confined in the negative tension brane (the visible
brane). Since $d=4$ gravity is localized on the hidden
brane, matter fields on the visible brane naturally have 
very weak gravitational coupling, so a large disparity
between  $M_{weak}$ and $M_{Planck}$. 

Attempts have been made to incorporate small neutrino mass
in the RS model \cite{grossman}. The model contains a bulk fermion
$\Psi$ and a bulk real scalar $\Phi$ with the following orbifold
boundary condition:
\beq
\Psi(-y)=\gamma_5\Psi(y), \quad
\Phi(-y)=-\Phi(y),
\eeq
in addition to the bulk gravition and the standard model
fields.
The action is given by
\bea
\label{rs}
S&=&\int d^4xdy \sqrt{-g_{(5)}}  [\, \frac{1}{2}M_*^3 R_{(5)}
-\Lambda_B
-\frac{i}{2}\bar{\Psi}\gamma^AD_A\Psi
-f\Phi\bar{\Psi}\Psi-m\bar{\Psi}^C\Psi+...]
\nonumber \\
&& +\int_{y=b} d^4x \sqrt{-g_{(4)}} [-\Lambda_v
+\kappa H\bar{\Psi}L+\kappa^{\prime}H\bar{\Psi}^CL+\frac{h}{M_*}LLHH+...]
\nonumber \\
&& +
+\int_{y=0} d^4x \sqrt{-g_{(4)}} [-\Lambda_h+...],
\eea
where $M_*$ is the 5-dimensional Planck mass,
$R_{(5)}$ is the $d=5$ Ricci scalar,
$\Psi^C$ is the charge-conjugation of the $d=5$ 
spinor $\Psi$, and $H$ and $L$ are the Higgs and lepton
doublets confined in the visible brane.
Here $\Lambda_v$ and $\Lambda_h$ denote the visible brane
tension and the hidden brane tension, respectively,
and we write explicitly also the terms for neutrino mass
in the visible brane action.

If the bulk and brane cosmological constants satisfy
\beq
k\equiv \sqrt{\frac{-\Lambda_B}{6M_*^3}}=\frac{\Lambda_h}{6M_*^3}
=-\frac{\Lambda_v}{6M_*^3}
\eeq
the model admits the following form of
$d=4$ Poincare invariant spacetime and the corresponding
massless $d=4$ graviton mode: 
\beq
ds^2=e^{-2k|y|}(n_{\mu\nu}+h_{\mu\nu}(x))dx^{\mu}dx^{\nu}-dy^2.
\quad (0\leq y \leq b).
\eeq 
Obviously $h_{\mu\nu}$ is localized around $y=0$, so
its coupling to the energy momentum tensor at $y=b$ 
is exponentially suppressed by $e^{-2kb}$.
Equivalently,
all dimensionful quantities on the visible brane
are rescaled by $e^{-2kb}$. This results
in the standard model mass parameter $M^2_{weak}\sim
e^{-2kb} M_*^2$, while the $d=4$ Planck mass is given by
$M^2_{Planck}=M_*^3(1-e^{-kb})/k$, so an exponentially small ratio
\beq
\frac{M_{weak}}{M_{Planck}}\sim e^{-kb}.
\eeq

The small ratio $m_{\nu}/M_{weak}$ can be similarly
obtained in the model of (\ref{rs}) by localizing 
the zero mode of $\Psi$ on the hidden brane.
To implement this mechanism, we need first the lepton
number violating couplings (both in bulk and on branes)
to be suppressed enough, for instance
$h$, $\kappa^{\prime}$ and $m/M_*$ should be less than $10^{-12}$
in order for $m_{\nu}\lesssim 0.1$ {\rm eV}.
This can be easily achieved by imposing a discrete
symmetry under which
\beq
\Psi\rightarrow e^{2\pi i/N}\Psi,
\quad
L\rightarrow e^{2\pi i/N} L,
\eeq
which would result in
\beq
h=\kappa^{\prime}=m=0.
\eeq
Then the Dirac equation for the zero mode of $\Psi$ is given by
\beq
\left(\frac{\partial}{\partial y}-2k+i\gamma_5f\langle\Phi\rangle
\right)\Psi_0=0,
\eeq
leading to the following solution
\beq
e^{-3k|y|/2}\Psi_0 = e^{-(2f\langle\Phi\rangle-k)|y|/2}\eta(x)
\eeq
where $\eta$ denotes the canonically normalized (in $d=4$ sense)
singlet neutrino mode.
On the parameter region with $k< 2f\langle \Phi\rangle$,
this mode is localized on the hidden brane.
As a result, $\eta$ has an exponentially small Yukawa coupling with 
$H$ and $L$ on the visible brane, so an
exponentially small Dirac neutrino mass.
After the proper rescaling of the involved fields,
one finds 
\beq
\frac{m_{\nu}}{M_{weak}}\sim
\kappa e^{-(2f\langle\Phi\rangle-k)b/2}
\eeq
which can be small as $10^{-12}$  to provide
the atmospheric neutrino mass.
Note that the small neutrino mass obtained by
localizing singlet neutrino on the hidden brane
is a Dirac mass, however
the current neutrino oscillation experiments
do not distinguish the Dirac mass from the Majorana mass.

\section{Models with abelian flavor symmetry}

Here we discuss some models in which the hierarchical patterns of 
the atmospheric and solar neutrino masses and mixing angles
are obtained by means of $U(1)$ flavor symmetries.
Our discussion will be limited to a specific example
for bi-maximal mixing with LMA MSW solar
neutrino oscillation and another example for
near-maximal atmospheric neutrino oscillation and SMA MSW
solar neutrino oscillation.

\medskip

{\bf A. A model for bi-maximal mixing}

\medskip

The neutrino masses and mixing angles for
bi-maximal mixing with LMA MSW solar neutrino
oscillation are given by
\bea
\label{lma}
&& m_2/m_3 \, \sim \, \lambda \, \, {\rm or} \, \, \lambda^2,
\nonumber \\
&& (\, |s_{23}|, \, |s_{12}|, \, |s_{13}|\,)
\, \sim \, (\,\frac{1}{\sqrt{2}},
\frac{1}{\sqrt{2}}, \lambda^k\,) \quad (k\geq 1).
\eea
One issue for this pattern of neutrino masses and mixing
angles is how could one  obtain small $\theta_{13}$ and
$m_2/m_3$  while keeping
$\theta_{23}$ and $\theta_{12}$ near maximal. 
Comparing the MNS mixing matrix $U=U^{e\dagger}U^{\nu}$
with the parametrization (\ref{para}),
one easily finds that $U$ automatically has a small
$\theta_{13}$ with bi-maximal $\theta_{23}$ and $\theta_{12}$ 
if $U^e$ has only one large mixing by $\theta_{23}$ and
also $U^\nu$ has only one large mixing by $\theta_{12}$.
A form of charged lepton mass matrix which would lead to
such $U^e$ is \beq
M^e \simeq m_\tau \pmatrix{ & & a\cr & & 1 \cr & & 1}
\label{emass}
\eeq
where $a \ll 1$. 
For the neutrino mass matrix, we can consider
two different forms leading to such $U^{\nu}$.
One is of pseudo-Dirac type:
\beq
\label{numass1}
M^{\nu} =
m_{3} \pmatrix{ b_1 & a & c\cr a & b_2 & d \cr c & d & 1}
\eeq
with $b_i\ll a$ and $c,d\ll 1$, and
the other is the plain large mixing between the 1st and 2nd
generations:
\beq
\label{numass2}
M^{\nu}=m_{3} \pmatrix{ a_1 & a_2 & c\cr a_2 & a_3 & d \cr c & d & 1},
\eeq
where all $a_i$ are of the same order and $c,d\ll 1$.
The neutrino mass  matrix is further constrained to
reproduce the correct mass ratio
$m_2/m_3\sim \lambda$ or $\lambda^2$.
Here we assume that $M^{\nu}$ is induced by
the conventional seesaw mechanism in supersymmetric model
and explore the possibility that the above mass matrix textures
are obtained as a consequence of 
$U(1)$ flavor symmetries.

It is not so trivial to obtain the mass matrix 
textures (\ref{emass}), (\ref{numass1}),
(\ref{numass2}) from $U(1)$ flavor
symmetries since $M^{\nu}$ needs to have
same order of magnitudes for the 1st and 2nd generations
while $M^e$ needs different ones.
This difficulty becomes more severe 
if we want to obtain a smaller value of $\theta_{13}$.
Among the models of $U(1)$ flavor symmetries,
the simplest one would be the case of single anomalous
$U(1)$ whose breaking is described by a single order
parameter $\epsilon=\langle \phi\rangle/M_*\sim \lambda$. 
Unfortunately, it turns out that the desired textures 
can not be obtained in this simplest case.
The next simple model  would be the case of
single non-anomalous $U(1)$ which has two symmetry breaking parameters 
with opposite $U(1)$ charges:
$$\epsilon=\frac{\langle\phi_+\rangle}{M_*},
\quad
\bar{\epsilon}=\frac{\langle \phi_-\rangle}{M_*},
$$
where $\phi_{\pm}$ has the $U(1)$ charge $\pm 1$.
Note that if the scale of $U(1)$ breaking is much higher
than the scale of supersymmetry breaking,
vanishing $U(1)$ $D$-term assures $|\epsilon|=|\bar{\epsilon}|$.
One can also consider the case of two $U(1)$'s  in which one
$U(1)$ is anomalous while the other is non-anomalous.
One plausible symmetry breaking pattern in this case is that
$\epsilon$ and $\bar{\epsilon}$ have the $U(1)\times U(1)$
charges $(-1,-1)$ and $(0,1)$, respectively.
In this case, vanishing  $D$-term of anomalous $U(1)$
leads to $|\epsilon|\sim \lambda$, while that of non-anomalous
$U(1)$ gives $|\epsilon|=|\bar{\epsilon}|$.
In the below, we will present a simple example for 
each case which gives rise to the mass matrix textures
for (\ref{lma}).

Let us first consider the case of single $U(1)$ with
two symmetry breaking parameters $\epsilon$ and $\bar{\epsilon}$.
We will assume that $|\epsilon|=|\bar{\epsilon}|\sim
\lambda$.
The light neutrino mass matrix which is obtained by the seesaw
mechanism is given by
\beq
M^{\nu} ={M^D (M^M)^{-1} (M^D)^T },
\eeq
where  $M^D$ is the $3\times 3$ Dirac mass matrix
 and $M^M$ is the $3\times 3$ Majorana mass matrix
of superheavy singlet neutrinos.
Let small letters denote the $U(1)$ charges of
the capital lettered superfields, e.g. $l_i$ for the
lepton doublets $L_i$, $e_i$ for the anti-lepton singlets
$E^c_i$, $n_i$ for the superheavy singlet neutrinos
$N_i$. Then  for the charge assignments of
\beq
l_i = (2,-2,0),\quad
n_i = (2,-2,0),\quad
e_i=(1,5,5),\quad h_1=h_2=0,
\eeq
one finds the mass matrices \cite{nir} which are are given by 
\beq
\label{massmatrix}
M^{\nu}=m_3\pmatrix{\lambda^4 & A & \lambda^2 \cr
             A & \lambda^4 & \lambda^2 \cr
             \lambda^2 & \lambda^2 & 1 }, \quad\quad
M^e=\langle H_1 \rangle \pmatrix{\lambda^7 & \lambda^7 & \lambda^3 \cr
             \lambda^3 & \lambda^3 & \lambda \cr
             \lambda^5 & \lambda^5 & \lambda }
\eeq
where $A$ is of order one, but does not exceed 1.
This  mass matrix  provides the near bi-maximal
$\theta_{23}$ and $\theta_{12}$, and also
\beq
\theta_{13}\sim \lambda^2, \quad 
\Delta m^2_{\rm atm}\sim (1-A^2)m_3^2,
\quad
\Delta m^2_{\rm sol}\sim 
\lambda^4 A^2 m_3^2,
\eeq
which can accomodate all experimental data
for reasonable values of $A$ and $m_3$.

The desired forms of mass matrices can be obtained
for the case of $U(1)\times U(1)$ also.
If one $U(1)$ is anomalous while
the other is non-anomalous,  which is the case
that appears quite often in compactified
string theory, it is quite plausible that
$U(1)\times U(1)$ are broken by the two symmetry breaking parameters
$\epsilon$ and $\bar{\epsilon}$ with
the $U(1)\times U(1)$ charges $(-1,-1)$ and $(0,1)$.
In this case,
$|\epsilon|=|\bar{\epsilon}|$ and they are naturally of
order the Cabbibo angle $\lambda$.
Then for the charge assignment
\bea
&& n_1=(0,-1), \quad
n_2=(0,1), \quad
n_3=(0,0),
\nonumber \\
&& l_1=(0,-1),
\quad
l_2=(1,2),
\quad
l_3=(0,0),
\nonumber 
\eea
one easily finds 
\beq
M^{\nu}\sim m_3\pmatrix{\lambda^2 & \lambda & \lambda \cr
\lambda & 0 & 0 \cr
\lambda & 0 & 1 }
\nonumber 
\eeq
which yields
\beq
\theta_{13}\, \sim \, \lambda,
\quad
\Delta m^2_{\rm sol}\, \sim \, \lambda^2 \Delta m^2_{\rm atm}.
\nonumber
\eeq

\medskip

{\bf B. A model for large atmospheric and small solar neutrino mixings}

\medskip

The neutrino masses and mixing angles for large atmospheric and small
solar neutrino mixings are given by
\bea
&& m_2/m_3 \, \sim \, \lambda^2, 
\nonumber  \\
&& (\, |s_{23}|, |s_{12}|, |s_{13}|\,)\, \sim \, (\,
\frac{1}{\sqrt{2}},\lambda^2,\lambda^k\,) \quad \quad (k\geq 1).
\eea
The issues for this pattern would be how could one obtain
small $m_2/m_3$ even when $\theta_{23}$ is near maximal, and also
what would be the reason for small $\theta_{12}$ and $\theta_{13}$.
Here we present a supersymmetric model with
$U(1)$ flavor symmetry in which such pattern of neutrino masses and 
mixing angles arises naturally.

The model under consideration is the MSSM with $R$-parity breaking
couplings which are suppressed by an anomalous $U(1)$ flavor symmetry
with $\epsilon\sim \lambda$ \cite{choi3}.
The most general $SU(3)_c\times SU(2)_L\times
U(1)_Y$-invariant superpotential of the MSSM superfields
includes the following lepton number ($L$) and $R$-parity violating
terms:
\beq
  \Delta W = \lambda_{ijk} L_i L_j E^c_k+\lambda^{\prime}_{ijk}
L_iQ_jD^c_k
\label{mssm}
\eeq
where  $(L_i, E^c_i)$ and $(Q_i, U^c_i, D^c_i)$  denote the lepton
and the quark superfields, respectively. 
Another $L$ and $R$-parity violating term $\mu^{\prime}_iL_iH_2$
in the superpotential can always be rotated away by
a unitary rotation of superfields.
Soft SUSY breaking terms also contain the $L$ and $R$-parity
violating terms:
\beq
\Delta V_{\rm soft}= m^2_{L_iH_1}L_iH_1^*+ B_iL_iH_2 
       + C_{ijk} L_i L_j E^c_k+C^{\prime}_{ijk}L_iQ_jD^c_k,
\eeq
where now all field variables denote the scalar components of the 
corresponding superfields.
In the basis in which $\mu_iL_iH_2$ in the
superpotential are rotated away, non-vanishing $B_i$ and
$m^2_{L_iH_1}$ result in the tree-level neutrino mass \cite{HS}
\begin{equation} \label{mntree}
 (M^{\nu})^{\rm tree}_{ij} \ap
% \frac{1}{2}\left( \frac{g^2_1}{M_1} + \frac{g^2_2}{M_2}\right)
  \frac{g_a^2 \langle\tilde{\nu}^*_i\rangle \langle\tilde{\nu}^*_j\rangle} 
{M_{1/2}} \,,
\end{equation}
where $M_{1/2}$ denote the $SU(2)_L\times U(1)_Y$  gaugino masses and the 
sneutrino VEV's are given by  
\beq
\langle\tilde{\nu}^*_i\rangle \,\,\ap\, \frac{2M_Z
(m^2_{L_iH_1}\cos\beta +B_i\sin\beta)}{
        m^2_{\tilde l}+\frac{1}{2}M_Z^2\cos 2\beta},
\eeq 
where $\tan\beta=\langle H_2\rangle/\langle H_1\rangle$,
$M_Z$ is the $Z$-boson mass, and
$m_{\tilde{l}}$ is the slepton soft mass
which is assumed to be (approximately) flavor-independent.
There are also additional neutrino masses arising from
various one-loop graphs involving the squark or slepton exchange
\cite{loopmass}.

Let  the small letters $q_i, u_i,$ e.t.c. denote the $U(1)$
charges of the superfields $Q_i, U^c_i,$ e.t.c.
If all $L$ and $R$-parity violating couplings
are suppressed by some powers of 
$\lambda$ as is determined by the $U(1)$ charges
of the corresponding operators,
the resulting  neutrino mass matrix takes the form:
\beq
(M^{\nu})_{ij}=m_3 \pmatrix{\lambda^{2l_{13}}A_{11} & \lambda^{l_{13}+l_{23}}
A_{12}
& \lambda^{l_{13}}A_{13} \cr
\lambda^{l_{13}+l_{23}}A_{12}
& \lambda^{2l_{23}}A_{22} &
\lambda^{l_{23}}A_{23} \cr
\lambda^{l_{13}}A_{13} & \lambda^{l_{23}}A_{23}
& A_{33}},
\label{mmat}
\eeq
where $m_3$ is the largest mass eigenvalue,
 $l_{i3}=l_i-l_3$, and all $A_{ij}$ are of order unity.
It is then straightforward to see that
the near maximal $\theta_{23}$ 
requires $l_2=l_3$, while the small $\theta_{12}\sim \lambda^2$
requires $l_1=l_2+2$.
This eventually leads to
the MNS  mixing matrix of the form:
\begin{equation} \label{}
U \sim \left( \begin{array}{ccc} 1 & \lambda^{2} & \lambda^{2} \\
                                       \lambda^{2} & 1 & 1 \\
                                       \lambda^{2} & 1 & 1 
                     \end{array} \right) \, . 
\end{equation}
which gives $\theta_{13}\sim \lambda^2$.

So far, we could get the mixing angle pattern
$\theta_{23}\sim 1$, $\theta_{12}\sim \theta_{13}\sim
\lambda^2$ just by assuming the $U(1)$ charge
relations:
$$
l_1-2=l_2=l_3.
$$
Still we need to get the mass hierarchies
$m_2/m_3\sim 4\times 10^{-2}$ and also $m_3/M_{weak}\sim
10^{-12}$.
In the model under consideration, $U(1)$ flavor symmetry
assures that $R$-parity violating couplings are all suppressed
by $\lambda^{l_i+h_2}$ or $\lambda^{l_i-h_1}$
compared to their $R$-parity conserving counterparts.
As a result, $m_3$  from $R$-parity violation
obeys roughly
$$
m_3/M_{weak} \, \sim \, \lambda^{2(l_3+h_2)} 
\quad {\rm or} \quad \lambda^{2(l_3-h_1)}.
$$ 
So if the $U(1)$ charges are arranged to have
$$l_3+h_2=l_3-h_1=7
\quad  {\rm or} \quad  8,
$$ the resulting $m_3$ naturally fits
into the atmospheric neutrino mass scale.

One may wonder how $m_2/m_3$ can be as small as $4\times 10^{-2}$
even when the 2nd and 3rd neutrinos mix maximally.
The neutrino mass matrix  (\ref{mmat})  from $R$-parity violation
automatically realizes such unusual scenario
since it is dominated by the tree 
mass (\ref{mntree}) which is a rank one matrix.
Then the largest mass $m_3$ is from
the tree contribution  while  $m_2$ is from the loops, so
$$
m_2/m_3 \, \sim \, {\rm LOOP}/{\rm TREE}
$$
{\it independently of} the value of $\theta_{23}$.
In fact, we need to make this loop to tree ratio a bit bigger
than the generic value in order to get the correct 
mass ratio $m_2/m_3\ap 4\times 10^{-2}$.
This is difficult to be achieved within the framework
of high scale SUSY breaking, e.g. gravity-mediated SUSY breaking
models, while it can be easily done
in gauge-mediated SUSY breaking models with
relatively low messenger scale.
In gauge-mediated SUSY breaking models \cite{gauge},
$B_i$ and $m^2_{L_iH_1}$ can be simultaneously rotated 
away as $\mu^{\prime}_i$ at the messenger scale $M_m$,
i.e. $B_i(M_m)=m^2_{L_iH_1}(M_m)=0$ in the basis
of $\mu^{\prime}_i=0$,  
and their low energy values at $M_{weak}$ are determined 
by the RG evolution.
Then the tree mass (\ref{mntree}) which is determined by
these RG-induced $B_i$ and $m^2_{L_iH_1}$ can be made smaller
by taking a lower value of the messenger scale.

Soft parameters  in  gauge-mediated models \cite{gauge} typically satisfy:
$M_a/\alpha_a\ap m_{\tilde q}/\alpha_3\ap
m_{\tilde l}/\alpha_{1,2}$ at $M_m$
where $M_a$, $m_{\tilde q}$, and $m_{\tilde l}$  denote the gaugino,
squark and slepton masses, respectively,
and $\alpha_a=g_a^2/4\pi$
for the standard model gauge coupling constants.
The size of the bilinear term $BH_1H_2$ in the scalar potential
depends upon how $\mu$ is generated. 
An attractive possibility 
is $B(M_m)=0$ for which
all CP-violating phases in soft parameters at $M_{weak}$ are automatically
small enough to avoid a too large electric dipole moment \cite{RS}.
In this case,
the RG-induced low energy value of $B$ yields
a large $\tan\beta\ap (m_{H_1}^2+m_{H_2}^2+2\mu^2)/B(M_Z)=40\sim 60$.

Analyzing the neutrino masses from $R$-parity violating couplings
which are determined by the RG evolution with the boundary conditions that
trilinear soft scalar couplings, $B$,
$B_i$ and $m^2_{L_iH_1}$ are all vanishing  at $M_m$,
and also $M_a/\alpha_a\ap m_{\tilde{q}}/\alpha_3\ap m_{\tilde{l}}/\alpha_{1,2}$
at $M_m$,
one finds \cite{choi3}
\beq
(M^{\nu})^{\rm tree}_{ij}\approx 
10^{-1}t^4 a_ia_j 
\left(\mu^2M^2_Z \over m^3_{\tilde{l}}\right) \,,
\eeq
where $a_i=y_b\lambda^{\prime}_{i33}
\sim\lambda^{l_i-h_1}$ for the $b$-quark Yukawa coupling
$y_b$ and $t=
\ln (M_m/m_{\tilde l})/\ln (10^3)$.
The loop mass is given
by \cite{choi3}
\beq
(M^{\nu})^{\rm loop}_{ij}\ap 10^{-2}t^2
y_by_{\tau}^3\lambda^{\prime}_{333}(\delta_{i3}\lambda_{j33}
+\delta_{j3}\lambda_{i33}) 
\left( \mu^2 M_Z^2 \over m^3_{\tilde{l}}\right)  \,,
\eeq
where $y_{\tau}$ is the $\tau$-lepton Yukawa coupling
and the smaller contributions are  ignored.
These tree and loop masses then give
the following mass hierarchies:
\bea
&& m_3/M_{weak}\ap [U_{i3}(M^{\nu})^{\rm tree}_{ij}U_{j3}]
/M_{weak}\, \sim \,  10^{-1}
\lambda^{2(l_3-h_1)} \,, 
\nonumber \\
&& m_2/m_3\ap [U_{i2}(M^{\nu})^{\rm loop}_{ij}U_{j2}]/m_3 \, \sim 
\, ({\rm LOOP}/{\rm TREE}),
\nonumber \\
&& m_1/m_2\ap [U_{i1}(M^{\nu})^{\rm loop}_{ij}U_{j1}]/m_2 \,\sim
\, \lambda^4\,, 
\eea
where 
$({\rm LOOP}/{\rm TREE})=10^{-2}
(\lambda_{233}/\lambda^{\prime}_{233})
(\ln10^3 /\ln{\frac{M_m}{m_{\tilde{l}}}})^2$,
and we have used
$\tan\beta\sim 50$,
$m_{\tilde l}\ap 300\GeV$ and 
$\mu\ap 2m_{\tilde l}$ 
which has been suggested to be the best parameter range for 
correct electroweak symmetry breaking \cite{RS}.
To summarize, in this model, small $m_2/m_3$ is due to
the loop to tree mass ratio, while the other small mass ratios
$m_1/m_2$ and $m_3/M_{weak}$ 
are from $U(1)$ flavor symmetry.

\bigskip

{\bf Acknowledgments}:
I thank E. J. Chun and K. Hwang for useful discussions, and also
Y. Kim for drawing the figures.
This work is supported by the BK21 project of the
Ministry of Education, KRF Grant No. 2000-015-DP0080,
KOSEF Grant No. 2000-1-11100-001-1, 
and KOSEF through the CHEP of KNU.

\begin{figure}
\centering
\epsfig{figure=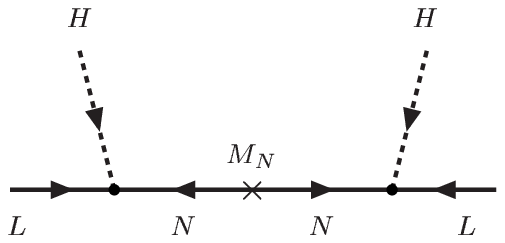,width=8cm}
\caption{Small neutrino mass from the exchange of superheavy singlet neutrino}
\end{figure}

\begin{figure}
\centering
\epsfig{figure=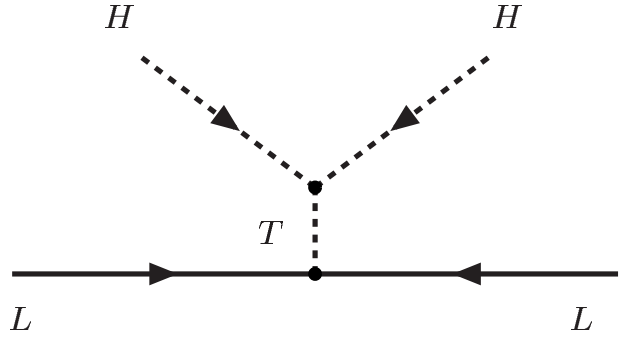,width=8cm}
\caption{Small neutrino mass from the exchange of superheavy triplet Higgs}
\end{figure}
\begin{figure}
\centering
\epsfig{figure=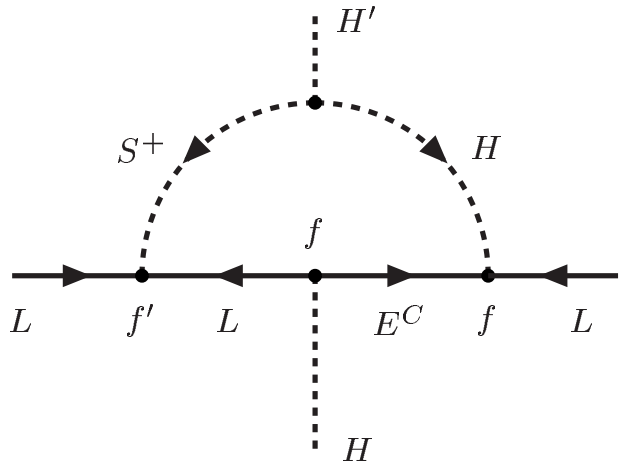,width=8cm}
\caption{One-loop neutrino mass in Zee-type  model}
\end{figure}
\begin{figure}
\centering
\epsfig{figure=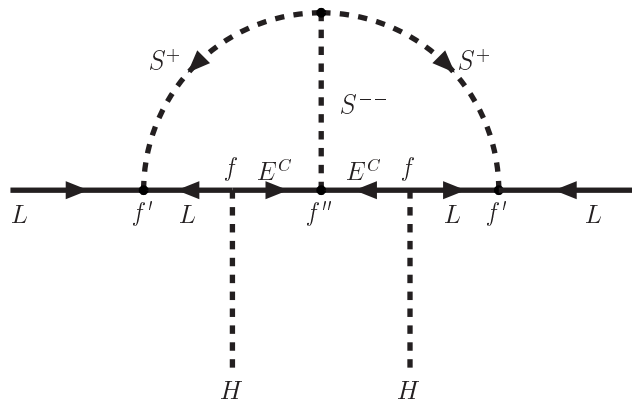,width=8cm}
\caption{Two-loop neutrino mass in the variant of Zee model}
\end{figure}

\end{document}